\documentclass[%
 reprint,
 superscriptaddress,
 showpacs,preprintnumbers,
 amsmath,amssymb,
 aps,
 pre,
 floatfix,
]{revtex4-1}

\usepackage{graphicx}% Include figure files
\usepackage{dcolumn}% Align table columns on decimal point
\usepackage{bm}% bold math
\usepackage{cancel}
\usepackage{color}
\usepackage[utf8]{inputenc}

\begin{document}

%\preprint{APS/123-QED}

%\title{On quantum statistics of classical particles:\\ a candidate for black energy}

\title{Quantum statistics of classical particles derived from\\ the condition of free diffusion coefficient}

\author{M. Hoyuelos}
\affiliation{Instituto de Investigaciones F\'isicas de Mar del Plata (IFIMAR -- CONICET), Funes 3350, 7600 Mar del Plata, Argentina}
\affiliation{Departamento de F\'isica, Facultad de Ciencias Exactas y Naturales, Universidad Nacional de Mar del Plata, Funes 3350, 7600 Mar del Plata, Argentina}

\author{P. Sisterna}
\affiliation{Departamento de F\'isica, Facultad de Ciencias Exactas y Naturales, Universidad Nacional de Mar del Plata, Funes 3350, 7600 Mar del Plata, Argentina}

\date{\today}% It is always \today, today,
             % but any date may be explicitly specified

\begin{abstract}
We derive an equation for the current of particles in energy space; particles are subject to a mean field effective potential that may represent quantum effects. From the assumption that non-interacting particles imply a free diffusion coefficient in energy space we derive Maxwell-Boltzmann, Fermi-Dirac and Bose-Einstein statistics. Other new statistics are associated to a free diffusion coefficient; their thermodynamic properties are analyzed using the grand partition function. A negative relation between pressure and energy density for low temperatures can be derived, suggesting a possible connection with cosmological dark energy models.
\end{abstract}


\maketitle

%##################################################################################
\section{Introduction}

Classical particles with the appropriate ---local--- interaction can have statistics typically associated to quantum systems. Kaniadakis and Quarati \cite{kania} introduced a nonlinear diffusion equation for the particle density in a classical system, whose study was extended in \cite{kania2} and \cite{kania3}. An appropriate choice of the transition probabilities, that depend on the particle density, gives rise to Fermi-Dirac or Bose-Einstein distributions in equilibrium. Using the analogy between quantum dynamics and classical dynamics with quantum fluctuations, the authors argue that the mentioned diffusion equation ``can describe a dynamics very close to the quantum dynamics of a system of identical particles.'' It was shown that this  model is useful to describe Bose-Einstein condensation \cite{toscani}. Gottesman \cite{gottesman} presented a model of classical bosons emphasizing the idea that indistinguishability is not an exclusive quantum property. The possibility of classical systems composed by indistinguishable particles is developed in \cite{saunders}.

A basic assumption in the derivation of quantum statistics is that the Hamiltonian has no interaction term. This is exactly the case for photons and approximately for electrons in a conductor.  We know that quantum effects can be reproduced in a classical system \textit{with} an interaction potential. For example, the Pauli exclusion principle for fermions is analogous to a potential that becomes infinite when two particles occupy the same state. This hard-core interaction is the cause of a nonlinear term in the corresponding diffusion equation, see \cite{kania}, \cite[p. 280]{frank} or Eq. (10) in \cite{suarez}. Nevertheless, we introduce the interaction only as a mean field effective potential in a Fokker-Planck equation.

In our classical system we consider that a particle can perform transitions to states that are close in energy. The transition probability from a state of energy $\epsilon$ to  a state of energy $\epsilon+\delta\epsilon$ is given by $P_{\epsilon,\epsilon+\delta\epsilon}$. It depends on $\epsilon$ and on the mean field potential in the involved states. In a non-equilibrium situation particles diffuse among energy states. Consideration of the diffusion coefficient in this energy space is central for our purposes. A system of non-interacting particles is obtained in the limit of low concentration; they diffuse with a free diffusion coefficient $D_0$ (for free particles). We may ask what happens with the diffusion coefficient when the concentration is high. In general, it depends on the mean field potential, since the transition probabilities depend on it. The simplest situation is when the mean field potential is zero; in this case the diffusion coefficient is $D_0$ for any concentration. We wish to explore the possibility that the diffusion coefficient remains equal to $D_0$ for any concentration even when the mean field potential is not zero, and find the potentials that satisfy this condition.
 
We conjecture that a free diffusion coefficient $D_0$ is a macroscopic manifestation of a non-interacting particle system at a microscopic level. A zero mean field potential corresponds to classical non-interacting particles. We find that a free diffusion coefficient can be associated not only to that case, but also to quantum statistics of non-interacting particles that require a non-zero mean field potential.

In other words, we consider two description levels. In a microscopic level, the Hamiltonian has no interaction term; occupation number in energy space may be different from free particle statistics due to quantum effects. We do not enter into the details of the microscopic level but, instead, develop a classical description based on an effective mean field potential that acts on a particle due to the presence of all the others. We argue that the absence of an interaction term at the microscopic level is manifested by a free diffusion coefficient in energy space. In the following, we show that statistics for fermions, bosons and free classical particles can be derived from this assumption. As a bonus, two new statistics are also obtained. The corresponding grand partition function is calculated, so that thermodynamic properties of these new kind of ---hypothetical--- particles can be analyzed.

\section{Diffusion in energy space}

We call $n_\epsilon$ the number of particles at energy level $\epsilon$. In this section we consider a description based on the behavior of \textit{one} particle in a mean field potential. The particle's energy has two separate contributions: one of the energy level, given by $\epsilon$; the other given by the mean field potential $\phi_\epsilon$. The potential is a function only of the number of particles; we use $\phi_\epsilon$ as an abbreviation of $\phi(n_\epsilon)$. It is a \textit{local} interaction in the sense that it holds for particles in the same energy level.

We impose the detailed balance condition on the transition probabilities:
\begin{equation}
e^{-\beta({\epsilon}+\delta\epsilon + \phi_{\epsilon+\delta\epsilon})} P_{\epsilon+\delta\epsilon,\epsilon} = e^{-\beta( {\epsilon} + \phi_\epsilon)} P_{\epsilon,\epsilon+\delta\epsilon}
\end{equation}
where $\beta = 1/k_B T$ and $T$ is the temperature of a reservoir. This condition guaranties the evolution to equilibrium, but it is not enough to specify $P_{\epsilon,\epsilon+\delta\epsilon}$. Arrhenius equation motivates the assumption that the transition probabilities can be written as a combination of exponentials of the energies involved:
\begin{equation}
P_{\epsilon,\epsilon+\delta\epsilon} = P\,\exp\left[\beta\, (a\, \phi_\epsilon + b\, \phi_{\epsilon + \delta\epsilon} +c\,\delta\epsilon)\right],
\end{equation}
where $a$, $b$ and $c$ are constants and $P$ may depend on $\epsilon$, it can be shown that the combination that fulfills detailed balance is $b=a-1$ and $c=1/2$. For later convenience, we introduce a parameter $\theta=1-2a$, so that
\begin{equation}
P_{\epsilon,\epsilon+\delta\epsilon} = P\,\exp\left[- \beta \left( \frac{\theta+1}{2}  \phi_{\epsilon+\delta\epsilon} + \frac{\theta-1}{2} \phi_\epsilon + \frac{1}{2}\delta\epsilon \right) \right].
\label{transprob}
\end{equation}
The implicit assumption in \eqref{transprob} is that the energy barrier of the Arrhenius equation is a function of the energies of the initial and final states in which parameter $\theta$ is introduced; see \cite{suarez2}. There is a clear interpretation for the cases $\theta = -1$, $0$ and $1$. For $\theta = -1$, the transition probability depends on the origin level:
\begin{equation}
P_{\epsilon,\epsilon+\delta\epsilon} = P e^{-\beta(-\phi_\epsilon + \delta\epsilon/2)}\quad\quad (\theta=-1).
\end{equation}
For $\theta = 0$, the transition depends on the energy difference between target and origin levels:
\begin{equation}
P_{\epsilon,\epsilon+\delta\epsilon} = P e^{-\beta(\phi_{\epsilon+\delta\epsilon} - \phi_\epsilon + \delta\epsilon)/2}\quad\quad (\theta=0).
\end{equation}
And for $\theta=1$ the transition depends on the target level:
\begin{equation}
P_{\epsilon,\epsilon+\delta\epsilon} = P e^{-\beta(\phi_{\epsilon+\delta\epsilon} + \delta\epsilon/2)}\quad\quad (\theta=1).
\end{equation}
These are the three cases that we analyze here. They are the ones for which the grand partition function is calculated, as we show later.

The current $J$ between levels $\epsilon$ and $\epsilon+\delta\epsilon$ is
\begin{equation}
J = n_\epsilon\, P_{\epsilon,\epsilon+\delta\epsilon} - n_{\epsilon+\delta\epsilon} \, P_{\epsilon+\delta\epsilon,\epsilon}.
\label{current}
\end{equation}
We replace \eqref{transprob} in \eqref{current}; the reversed transition probability is obtained by exchanging $\epsilon \leftrightarrow \epsilon + \delta\epsilon$ in \eqref{transprob}. The approximations $(n_{\epsilon+\delta\epsilon} - n_\epsilon)/\delta\epsilon \simeq \frac{\partial n}{\partial \epsilon} $ and $(\phi_{\epsilon+\delta\epsilon}- \phi_\epsilon)/ \delta\epsilon \simeq \frac{d \phi}{d n} \frac{\partial n}{\partial \epsilon}$ are used. After some algebra (see the appendix for details), we obtain, in the continuous limit,
\begin{equation}
J = \left[-D_0\,e^{-\beta\theta \phi} \beta\, n - D_0\,e^{-\beta\theta \phi} \left( \beta n\frac{d\phi}{dn} + 1\right) \frac{\partial n}{\partial \epsilon} \right] \frac{1}{\delta\epsilon},
\label{ccurrent}
\end{equation}
where $D_0=P\,\delta\epsilon^2$ is the free diffusion coefficient. The evolution of the number of particles is given by the continuity equation in energy space: $\frac{1}{\delta\epsilon}\frac{\partial n}{\partial t} = - \frac{\partial J}{\partial \epsilon}$ (let us note that factor $1/\delta \epsilon$ here and in \eqref{ccurrent} vanishes when it is constant and equations are written in terms of the particle concentration per unit energy: $n/\delta\epsilon$). Identifying the zero current state with equilibrium, it is easy to see that the equilibrium number of particles is $n_\textrm{eq} \propto e^{-\beta(\phi + \epsilon)}$. The proportionality constant can be written as
\begin{equation}
n_\textrm{eq} = e^{-\beta(\phi + \epsilon - \mu)},
\label{neq}
\end{equation}
where we can identify $\mu$ with the chemical potential. The factor in front of $\frac{\partial n}{\partial \epsilon}$, in \eqref{ccurrent}, is the diffusion coefficient,
\begin{equation}
D = D_0\,e^{-\beta\theta \phi} \left( \beta n\frac{d\phi}{dn} + 1\right).
\label{difcoef}
\end{equation}

\section{Statistics for $D=D_0$}

The condition $D=D_0$ implies
\begin{equation}
\frac{d\phi}{dn} = \frac{e^{\beta\theta \phi}-1}{\beta\, n}.
\label{vinterac}
\end{equation}
The free particle potential, $\phi=0$, is always solution of \eqref{vinterac}.  It is the only solution for $\theta=0$. Replaced in \eqref{neq}, it gives the Maxwell-Boltzmann distribution:
\begin{equation}
n_\textrm{eq} = e^{-\beta(\epsilon - \mu)}\quad\quad(\theta=0).
\label{mbdist}
\end{equation}
For $\theta \neq 0$ there are other solutions:
\begin{equation}
\phi(n) = -\frac{1}{\theta \beta} \ln(1 - \kappa\,n^\theta).
\label{soltheta}
\end{equation}
%We recall that this effective potential does not imply interaction at a microscopic level; it is analogous to a statistical potential; see, for example, \cite[p. 138]{pathria}. 
Let us consider the case $\theta=1$. Replacing \eqref{soltheta} in \eqref{neq} we obtain
\begin{equation}
n_\textrm{eq} = \frac{1}{e^{\beta(\epsilon-\mu)} + \kappa } \quad\quad (\theta=1).
\label{fdbe}
\end{equation}
The values $\kappa$ equal to $1$ and $-1$ correspond to Fermi-Dirac and Bose-Einstein statistics respectively. The authors of \citep{kania} arrive to the same result using a different procedure, and propose the continuous variation of parameter $\kappa$ between $-1$ and 1 to produce intermediate statistics (see \cite{kana} for a review on mathematical aspects of parastatistics). The absolute value of $\kappa$ may be absorbed in the change of variables $n_\textrm{eq} \rightarrow n_\textrm{eq}/|\kappa|$, $\mu \rightarrow \mu - \beta^{-1} \ln|\kappa|$.

For $\theta=-1$ we obtain
\begin{equation}
n_\textrm{eq} = e^{-\beta(\epsilon-\mu)} + \kappa \quad\quad (\theta=-1).
\label{newstat}
\end{equation}
As for fermions and bosons, we will consider that only the sign of $\kappa$ is relevant, so it takes the values 1 or $-1$. This is a result that, as far as we know, was not previously reported. Since we arrived to it following the same procedure that took us to Fermi-Dirac, Bose-Einstein and Maxwell-Boltzmann distributions, we consider that a deeper analysis is worth it. As shown in the next section, we can get more information using the grand partition function.

The necessity of identifying the particles that obey \eqref{newstat} with specific names will arise in the next sections. We use the names \textit{ewkons} and \textit{genkons} (the elder and younger sister in tehuelche language) for the particles obeying \eqref{newstat} with $\kappa$ equal 1 and $-1$ respectively. There is a problem with genkons, since for large enough energy the equilibrium particle number, $n_\textrm{eq}$, becomes negative, and this implies a complex mean field potential. The proposition of a physical interpretation for genkons would be rather speculative at this stage. Therefore, we will focus our attention on the properties of ewkons.

\section{The grand partition function}

Let us consider a system with a discrete set of energy levels. The grand partition function for indistinguishable classical particles in level $\epsilon$ is:
\begin{equation}
\mathcal{Z}_{\epsilon, \textrm{cla}} =  \sum_{N=0}^{\infty} \frac{1}{N!} e^{-\beta(\epsilon - \mu) N} = \exp\left( e^{-\beta (\epsilon - \mu)}\right).
\end{equation}
The Maxwell-Boltzmann distribution \eqref{mbdist} is immediately obtained using $n_\textrm{eq} = \beta^{-1} \frac{\partial \ln \mathcal{Z_\epsilon}}{\partial \mu}$.

\begin{figure}
\includegraphics[scale=.7]{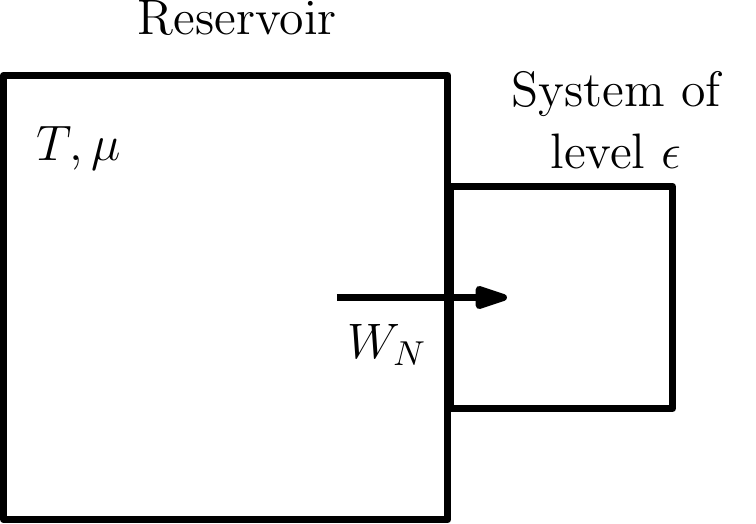}
\caption{Scheme of level $\epsilon$ system in contact with a reservoir with temperature $T$ and chemical potential $\mu$. The reservoir performs work $W_N$ on the system.}
\label{esq}
\end{figure}

Quantum effects can be introduced by considering that the system of level $\epsilon$ is in contact with a reservoir with which not only heat and particles are exchanged, but also work; see figure \ref{esq}. The elementary system in
the grand canonical ensemble consists of all the particles within a given energy level. The system's energy is, as always, $\epsilon N$, and the reservoir's energy change is $- \mu N - W_N$, where $W_N$ is an extra work done on the system, necessary to accommodate $N$ particles. (The work done by a reservoir on a system is treated in this way in, for example, the Gibbs canonical ensemble; see \cite[p. 115]{kardar}.) Therefore,
\begin{equation}
\mathcal{Z}_\epsilon =  \sum_{N=0}^{\infty} \frac{1}{N!} e^{-\beta(\epsilon N - \mu N - W_N)}.
\label{partfunc}
\end{equation}
This approach looks more involved than the usual one for deriving the grand partition function for bosons or fermions. But, as we show in the next paragraphs, it allows the derivation of an interesting relation between $W_N$ and the effective potential $\phi(n)$, with $n = \langle N \rangle$; we omit subscript `eq' in $n$ for simplicity. And, also, the procedure drives us to the new statistics for ewkons.

We connect both descriptions ---mean field and grand partition function--- by the condition that the equilibrium number of particles should be the same in both cases. Using \eqref{neq} we have
\begin{align}
e^{-\beta(\phi + \epsilon - \mu)} &= \frac{1}{\mathcal{Z}_\epsilon} \sum_{N=0}^\infty \frac{N}{N!} e^{-\beta(\epsilon N - \mu N - W_N)}
\end{align}
and, after some simplification,
\begin{align}
e^{-\beta \phi} &= \frac{1}{\mathcal{Z}_\epsilon} \sum_{N=1}^\infty \frac{1}{(N-1)!} e^{-\beta[\epsilon (N-1) - \mu (N-1) - W_N]} \nonumber \\
&= \frac{1}{\mathcal{Z}_\epsilon} \sum_{N'=0}^\infty \frac{1}{N'!} e^{-\beta[\epsilon N' - \mu N' - W_{N'}]} \; e^{-\beta(W_{N'} - W_{N'+1})} \nonumber \\
&= \left\langle e^{-\beta(W_{N} - W_{N+1})} \right\rangle. \label{phiW}
\end{align}
A physical interpretation is possible using the Jarzynski equality \cite{jarzynski}, also called Bochkov-Kuzovlev-Jarzynski equality
\cite{bochkov}. It gives a relation between the variation of the Helmholtz free energy $\Delta F$ and the work $W$,
\begin{equation}
e^{-\beta\, \Delta F} = \left\langle e^{-\beta\, W} \right\rangle.
\end{equation}
We can see that the effective potential $\phi(n)$ is equal to the free energy change when the work $W = W_{N} - W_{N+1}$ is applied.

Let us consider the case of fermions. The state given by two or more fermions in the same level is forbidden through a divergence of the energy of system and reservoir. The system's energy remains finite and equal to $\epsilon N$ due to the assumption of non-interacting particles. The divergence is carried out by the reservoir that performs a work on the system that tends to $-\infty$. We have:
\begin{equation}
W_{N, \textrm{fer}} = \left\{ \begin{array}{ll}
0 & \mbox{for } N = 0, 1 \\ 
-\infty & \mbox{for } N \ge 2
\end{array} \right.
\end{equation}
Replacing in \eqref{partfunc} we obtain the corresponding partition function 
$\mathcal{Z}_{\epsilon,\textrm{fer}} = 1 + e^{-\beta(\epsilon - \mu)}$. And replacing in \eqref{phiW} we get 
\begin{equation}
e^{-\beta \phi_\textrm{fer}} = \frac{1}{\mathcal{Z}_{\epsilon,\textrm{fer}}} = 1 - n,
\end{equation}
where we used the relation $n = 1/(1 + e^{\beta(\epsilon - \mu)})$ for fermions. Then, we recover equation \eqref{soltheta} with $\theta=1$ and $\kappa = 1$:
\begin{equation}
\phi_\textrm{fer}(n) = - \beta^{-1} \ln(1-n).
\end{equation}

The partition function for bosons, $\mathcal{Z}_{\epsilon,\textrm{bos}} =  (1-e^{-\beta(\epsilon - \mu)})^{-1}$, is obtained when the Gibbs factor $1/N!$ in \eqref{partfunc} is canceled with the corresponding work:
\begin{equation}
W_{N,\textrm{bos}} = \beta^{-1} \ln N!
\end{equation}
The effective mean field potential is derived using \eqref{phiW}:
\begin{equation}
e^{-\beta \phi_\textrm{bos}} = \langle 1 + N \rangle = 1 + n,
\end{equation}
with $n=1/(e^{\beta(\epsilon - \mu)}-1)$, expression that corresponds to \eqref{soltheta} with $\theta=1$ and $\kappa = -1$.

Once the procedure is established for the well known cases of fermions and bosons, we can move to the new statistics of ewkons. The effective potential for ewkons, equation \eqref{soltheta} with $\theta=-1$ and $\kappa = 1$, is
\begin{equation}
\phi_\textrm{ewk}(n) = \beta^{-1} \ln(1 - 1/n).
\end{equation}
Using \eqref{phiW}, we have
\begin{equation}
\frac{n}{n-1} = \left\langle e^{-\beta(W_{N,\textrm{ewk}} - W_{N+1,\textrm{ewk}})} \right\rangle.
\label{phiwewk}
\end{equation}
The problem to determine $W_{N,\textrm{ewk}}$ is that now we do not have the guide of a known partition function as in the previous cases. Nevertheless, we can prove that a simple and monotonic solution of \eqref{phiwewk} is given by
\begin{equation}
W_{N,\textrm{ewk}} = \beta^{-1} \ln N.
\label{Wewk}
\end{equation}
The corresponding grand partition function is  
\begin{align}
\mathcal{Z}_{\epsilon,\textrm{ewk}} &=  \sum_{N=0}^\infty \frac{1}{N!} e^{-N\beta(\epsilon-\mu) + \ln N} \nonumber \\
& =  \sum_{N=1}^\infty \frac{1}{(N-1)!} e^{-N\beta(\epsilon-\mu)} \nonumber \\
& =  \sum_{N'=0}^\infty \frac{1}{N'!} e^{-(N'+1)\beta(\epsilon-\mu)} \nonumber \\
& =  \exp\left[ e^{-\beta(\epsilon-\mu)} - \beta(\epsilon - \mu)\right]
\label{partewk}
\end{align}
The mean particle number 
\begin{equation}
n = \beta^{-1} \frac{\partial \ln \mathcal{Z}_{\epsilon,\textrm{ewk}}}{\partial \mu} = e^{-\beta (\epsilon - \mu)} +1
\label{newk}
\end{equation}
is equivalent to \eqref{newstat} with $\kappa = 1$. In order to check that expression for the work $W_{N,\textrm{ewk}}$ \eqref{Wewk} satisfies the relation \eqref{phiwewk} we make the replacement and obtain:
\begin{align}
\frac{n}{n-1} &= \left\langle \frac{N+1}{N} \right\rangle \nonumber \\
&= \frac{1}{\mathcal{Z}_{\epsilon,\textrm{ewk}}} \sum_{N=0}^\infty \frac{1}{N!} \left( 1 + \frac{1}{N}\right) e^{-N\beta(\epsilon-\mu) + \ln N} \nonumber \\
&= 1 + e^{\beta (\epsilon - \mu)}.
\end{align}
It can be seen that this equality is satisfied using the result for the mean particle number for ewkons \eqref{newk}.

\section{Thermodynamic properties}

The average energy $E_\epsilon$ of level $\epsilon$ can be evaluated with
\begin{equation}
E_\epsilon - \mu n = - \frac{\partial \ln\mathcal{Z}_\epsilon}{\partial \beta}.
\label{relatione}
\end{equation}
The same result is obtained for fermions, bosons or ewkons, as expected for non-interacting particles:
\begin{equation}
E_\epsilon = n\, \epsilon, \label{avenergy}
\end{equation}
To obtain the total energy we need the total grand partition function given by
\begin{equation}
\mathcal{Z} = \prod_{\{\epsilon\}} \mathcal{Z}_\epsilon,
\end{equation}
where the product is performed on all energy levels, repeated if necessary depending on the degeneracy. Let us consider the case of ewkons, we have:
\begin{align}
\ln \mathcal{Z} &= \sum_{\{\epsilon\}} \ln \mathcal{Z}_{\epsilon,\textrm{ewk}} \nonumber \\
&= \sum_{\{\epsilon\}} \left[ e^{-\beta(\epsilon-\mu)} - \beta(\epsilon - \mu)\right].
\end{align}

The usual approach to derive thermodynamic properties of an ideal quantum gas of bosons or fermions assumes a continuous energy spectrum and a density of states $g(\epsilon)$ proportional to $\epsilon^{1/2}$. If we try to reproduce the same steps for a gas of ewkons in a volume $V$ we obtain divergent values of energy or particle density at finite temperature. The divergences can be removed by an appropriate choice of the density of states, but we do not have arguments to support a specific one. Nevertheless, some interesting information can still be extracted form the grand partition function.

Assuming that the energy gaps are small compared to the average energy, the sum over states is transformed to an integral, and the total energy, total number of particles and pressure are written as:
\begin{align}
E &= \int d\epsilon \;g(\epsilon)\, \epsilon \,\left( e^{-\beta (\epsilon - \mu)} + 1 \right) \\
\mathcal{N} &= \int d\epsilon \; g(\epsilon) \left( e^{-\beta (\epsilon - \mu)} + 1 \right) \\
P &= \frac{1}{V \beta} \ln \mathcal{Z} \nonumber \\
&= \frac{1}{V \beta} \int d\epsilon \; g(\epsilon) \left( e^{-\beta (\epsilon - \mu)} - \beta (\epsilon-\mu) \right) 
\end{align}
The entropy is
\begin{align}
S &= k_B \frac{\partial (T \ln \mathcal{Z})}{\partial T} \nonumber \\
&= k_B \int d\epsilon \; g(\epsilon)\, e^{-\beta (\epsilon - \mu)} \left( 1 + \beta (\epsilon - \mu)\right).
\end{align}
We note that to keep the entropy ---and the other quantities--- bounded in the limit $\beta \rightarrow \infty$, we have to consider the condition $\epsilon_\textrm{min} > \mu$, where $\epsilon_\textrm{min}$ is the minimum value of the energy for which $g(\epsilon)\neq 0$. 

Let us consider the situation of small temperature and $\epsilon_\textrm{min} > \mu$. In this case we have
\begin{align}
E &\simeq   \int d\epsilon \;g(\epsilon)\, \epsilon \\
P V &\simeq - \int d\epsilon \;g(\epsilon) (\epsilon - \mu).
\end{align}
The relation between pressure and energy density is
\begin{equation}
w = \frac{PV}{E} = -1 + \frac{\int d\epsilon \;g(\epsilon)\, \mu}{\int d\epsilon \;g(\epsilon)\, \epsilon} < 0
\end{equation}
where the inequality comes from the condition $\epsilon_\textrm{min} > \mu$. This result is in contrast with the one corresponding to fermions or bosons. An ideal non-relativistic gas of fermions or bosons has an always positive relation between pressure and energy density, $w = 2/3$; see \cite[p. 189]{kardar}.

\section{Examples}

In this section we evaluate thermodynamic quantities of an ewkon gas at low temperature for several forms of the density of states $g(\epsilon)$. As mentioned before, we can not support a specific choice of $g(\epsilon)$. The only condition that we impose is that the thermodynamic quantities do not diverge. The examples are useful to obtain concrete values of these quantities, to check general properties and to have a flavor of what kind of densities are appropriate for ewkons (the same approach was implemented in other contexts; see, for example, \cite{matty,hilbert}). Divergences are avoided with fast decreasing densities, or with non-zero densities in a bounded region of energy.

\subsection{Exponentially decreasing density of states}

Let us consider that the density of states is given by
\begin{equation}
g(\epsilon) = \left\{ \begin{array}{cc}
B e^{-c \epsilon} & \mbox{if } \epsilon > \epsilon_0 \\ 
0 & \mbox{if } \epsilon < \epsilon_0
\end{array}  \right.
\end{equation}
with $\epsilon_0 \geq 0$; constants $B$ and $c$ have dimensions of the inverse of energy. The total energy, number of particles and pressure are 
\begin{align*}
E &= B e^{-c\epsilon_0} \left[e^{-\beta(\epsilon_0-\mu)}\frac{\epsilon_0(\beta+c)+1}{(\beta+c)^2} + \frac{\epsilon_0}{c} + \frac{1}{c^2} \right] \\
\mathcal{N} &= B e^{-c\epsilon_0} \left[ \frac{e^{-\beta(\epsilon_0-\mu)}}{\beta+c}  + \frac{1}{c}  \right] \\
P &= \frac{B e^{-c\epsilon_0}}{V} \left[ \frac{e^{-\beta(\epsilon_0-\mu)}}{\beta(\beta+c)} + \frac{\mu-\epsilon_0}{c} - \frac{1}{c^2} \right].
\end{align*}

In the limit of low temperature, and taking into account the condition $\epsilon_0 > \mu$, the relation between pressure and energy density is
\begin{equation}
w = -1 + \frac{c\mu}{1 + c \epsilon_0} < 0.
\end{equation}

At high temperature, $\beta \rightarrow 0$, the energy and total number of particles remain bounded:
\begin{align*}
E &\rightarrow B e^{-c\epsilon_0} 2 (\epsilon_0 + 1/c) \\
\mathcal{N} &\rightarrow B e^{-c\epsilon_0} 2/c,
\end{align*}
but the pressure diverges, and also $w$.

\subsection{Quadratic density of states}

We now consider a density of states that is different from zero in a bounded interval of energy and has a quadratic form:
\begin{equation}
g(\epsilon) = \frac{6 \Omega}{\epsilon_s^3}\epsilon \,(\epsilon_s-\epsilon)\,\Theta(\epsilon) \,\Theta(\epsilon_s-\epsilon)
\end{equation}
where $\Omega = \int d\epsilon \; g(\epsilon)$ is the total number of states, and $\Theta$ is the Heaviside step function. The density is different from zero in the interval $0 < \epsilon < \epsilon_s$; we have the condition $\mu<\epsilon_\textrm{min} = 0$. The energy and number of particles are
\begin{align*}
E =&\ \frac{6 \Omega}{\epsilon_s^3}\left[ \frac{e^{-\beta(\epsilon_s-\mu)}}{\beta^2}\left(\epsilon_s^2 +\frac{4\epsilon_s}{\beta} + \frac{6}{\beta^2}\right) \right. \\
& + \left. \frac{e^{\beta \mu}}{\beta^3}\left(2\epsilon_s-\frac{6}{\beta}\right) + \frac{\epsilon_s^4}{12}\right] \\
\mathcal{N} = &\ \Omega + \frac{6\Omega e^{\beta \mu}}{\beta^2 \epsilon_s^2}\left[ e^{-\beta\epsilon_s}\left(1+\frac{2}{\beta\epsilon_s}\right) + 1- \frac{2}{\beta\epsilon_s} \right].
\end{align*}
The pressure can be written in terms of the number of particles giving the equation of state
\begin{equation}
PV = N/\beta - \Omega\,(1/\beta - \mu) - \Omega \epsilon_s/2.
\end{equation}

The relation between pressure and energy density at low temperature is
\begin{equation}
w = -1 + \frac{2\mu}{\epsilon_s} < 0 \label{omegaquadratic}.
\end{equation}

In the limit of high temperature, all the quantities, $E$, $\mathcal{N}$, $P$ and $w$ diverge to $+\infty$.

\section{Conclusions}

The conjecture that a free particle diffusion coefficient in energy space is a characteristic of a system composed by non-interacting particles led us to the known particle distributions of Maxwell-Boltzmann, Fermi-Dirac and Bose-Einstein and to two unknown distributions of particles that we called ewkons and genkons. We focused our analysis on ewkons, since genkons can have a negative number of particles whose justification may require a development in a different context.

The free diffusion coefficient condition might be seen as a key idea for an alternative method to derive quantum statistics for bosons and fermions. Nevertheless, from equations \eqref{neq} and \eqref{soltheta} we obtain a family of particle distributions 
\begin{equation}
n = \left( e^{\theta \beta (\epsilon-\mu)} + \kappa\right)^{-1/\theta}
\end{equation}
from which we analyzed the cases $\theta = -1, 0, 1$.  We consider that it is interesting to analyze the properties of these new distributions from a theoretical point of view. But even more interesting would be to explore the possibility to find such distributions in nature. Nature phenomena are described in terms of only two kind of particles: fermions and bosons. However, room for speculation grew up since the discovery of dark energy and dark matter, corresponding to $95\%$ of the total amount of matter of the universe. Many candidates have been proposed for the constituent of this dark energy and matter, but their nature remains elusive. The low pressure and low energy density of the universe suggest the assumption of an ideal gas in the formulation of a cosmological equation of state. As mentioned before, an ideal Fermi or Bose gas has a positive parameter $w$. But the observed accelerated expansion of the universe \cite{hogan,Planck2015} implies a negative $w$. For instance, using early dark energy parameterizations (which encompass features of a large class of dynamical dark energy models), Planck collaboration latest release implies that $w_0<-0.93$ with a 95\% confidence level, where $w_0$ is the present value of  $w$. Another interesting case is the dark energy coupled scenarios \cite{Amendola}, where there is a fifth force between dark matter particles mediated by the dark energy scalar field. In this case the results of \cite{Planck2015} show that $w_0$ does not differ from $-1$ by more that $1\%$. The result $w<0$ for a gas of ewkons at low temperature is a promising starting point for a candidate to dark energy or matter. A question for the future would be: is there any dark energy model such that the quantum field theory involved has particles with the statistics of ewkons? Generic models of slow roll dark energy such as those studied in \cite{Gott} has varying $w$ given by $\delta w(z)\simeq\delta w_0 (H_o/H(z))^2$, where $H_0$ and $H(z)$ are the Hubble constant at present and at red-shift $z$ respectively, and $\delta w\equiv w+1$. Anyway, the conservative model-independent bound $\delta w_0<0.1$ implies for ewkons with quadratic density of states the upper bound $\mu/\epsilon_s<0.05$.

There are several open problems that deserve further analysis in the future: a more general study of particle distributions in terms of $\theta$, the search for a justification of the possible specific shape of the density of states, and the physical interpretation of a negative number of particles or a negative total energy for genkons. We know that quantum effects related to the parity of a quantum state give rise to fermion and boson statistics, so it would be very important to determine what are the corresponding effects at the microscopic level that give rise to ewkon or genkon statistics. Taking for instance the model we studied with quadratic density of states we may ask: could there be dark energy with a chemical potential (see \eqref{omegaquadratic}) such that $\mu(z)$ has a similar cosmological dependence as slow-roll dark energy requires? Mass-varying neutrinos are another candidate for dark energy \cite{Caldwell}. Could their statistics show ewkon like features? We look forward to answer some of these questions in the future.      

\section*{Appendix}

More details for the derivation of the current in energy space \eqref{ccurrent} are presented in this appendix. Replacing the expression for the transition probabilities \eqref{transprob} in \eqref{current}, we have
\begin{align*}
J =&\ n_\epsilon P e^{-\beta\left[ (\theta + 1) \phi_{\epsilon+\delta\epsilon} + (\theta-1)\phi_\epsilon + \delta\epsilon \right]/2} \\
& - n_{\epsilon+\delta\epsilon} P e^{-\beta\left[ (\theta + 1)\phi_{\epsilon} + (\theta-1) \phi_{\epsilon+\delta\epsilon} - \delta\epsilon \right]/2}
\end{align*}
In the continuous limit the number of particles for discrete values of the energy, $n_\epsilon$, is replaced by a function $n$ of $\epsilon$, and the mean field potential $\phi_\epsilon$ by $\phi$, a function of $n$. Not only the discrete energy space becomes continuous, also the number of particles, that now represents an average over samples; correlations in non-linear terms are neglected when taking the average of the previous equation (Ginzburg criterion). The  following approximations are used: $n_{\epsilon+\delta\epsilon} \simeq n + \frac{\partial n}{\partial \epsilon} \delta\epsilon$ and $\phi_{\epsilon+\delta\epsilon} \simeq \phi + \frac{d \phi}{d n} \frac{\partial n}{\partial \epsilon} \delta\epsilon$. Replacing in the previous expression, we have
\begin{align*}
J& = P e^{-\beta\theta\phi} \left\{ n \exp\left[-\beta \left((\theta+1)\frac{d\phi}{dn}\frac{\partial n}{\partial \epsilon} + 1\right)\frac{\delta\epsilon}{2} \right] \right. \\
& - \left.\left(n+\frac{\partial n}{\partial \epsilon}\delta\epsilon\right)\,\exp\left[-\beta \left((\theta-1)\frac{d\phi}{dn}\frac{\partial n}{\partial\epsilon} - 1\right)\frac{\delta\epsilon}{2}\right]\right\}.
\end{align*}
The exponentials inside the brackets are expanded up to order $\delta\epsilon$:
\begin{align*}
J& = P e^{-\beta\theta\phi} \left\{ n \left[1-\beta \left((\theta+1)\frac{d\phi}{dn}\frac{\partial n}{\partial \epsilon} + 1\right)\frac{\delta\epsilon}{2} \right] \right. \\
& - \left.\left(n+\frac{\partial n}{\partial \epsilon}\delta\epsilon\right)\,\left[1-\beta \left((\theta-1)\frac{d\phi}{dn}\frac{\partial n}{\partial\epsilon} - 1\right)\frac{\delta\epsilon}{2}\right]\right\}
 \\
&= -D_0\, e^{-\beta\theta\phi} \left\{\beta n + \left(\beta n \frac{d\phi}{dn} + 1\right) \frac{\partial n}{\partial \epsilon} \right\}\frac{1}{\delta\epsilon},
\end{align*}
where the free diffusion coefficient $D_0= P\,\delta\epsilon^2$ was introduced.
The factor multiplying $\frac{\partial n}{\partial \epsilon}$ is the diffusion coefficient \eqref{difcoef}. The other term, proportional to $n$, is a drift term that drives the particles towards low energy states.

\begin{acknowledgments}
M. H. acknowledges help provided by useful discussions with H. O. Mártin during preliminary stages of this work.	
\end{acknowledgments}

\end{document}